\documentclass[10pt,numbers,reprint]{sigplanconf}

\pdfoutput=1

\newcommand{\ignore}[1]{}

\usepackage{fancyhdr}
\usepackage[normalem]{ulem}
\usepackage[hyphens]{url}
\usepackage{hyperref}

\usepackage{amssymb}
\usepackage{amsmath}
\usepackage{graphicx}

\usepackage{balance}

\usepackage{booktabs}   
\usepackage{subcaption} 
\usepackage{tikz}
\usepackage{pgfplots}

\title{{BOLT}: {A} Practical Binary Optimizer for Data Centers and Beyond}

\authorinfo{Maksim Panchenko \and Rafael Auler \and Bill Nell \and Guilherme Ottoni}
           {Facebook, Inc.\\Menlo Park, CA, USA}
           {\{maks,rafaelauler,bnell,ottoni\}@fb.com}

\begin{document}
\makeatletter
\def\@copyrightspace{\relax}
\makeatother

\maketitle




\begin{abstract}
Performance optimization for large-scale applications has recently
become more important as computation continues to move towards
data centers.  Data-center applications are generally very large and
complex, which makes code layout an important optimization to improve
their performance.  This has motivated recent investigation of
practical techniques to improve code layout at both compile time and
link time.  Although post-link optimizers had some success in the
past, no recent work has explored their benefits 
in the context of modern data-center applications.

In this paper, we present BOLT, a post-link optimizer built on top of
the LLVM framework. Utilizing sample-based profiling, BOLT boosts
the performance of real-world applications even for 
highly optimized binaries built with both
feedback-driven optimizations~(FDO) and link-time optimizations~(LTO).
We demonstrate that post-link performance improvements are complementary
to conventional compiler optimizations, even when the latter are done at
a whole-program level and in the presence of profile information.
We evaluated BOLT on both Facebook data-center workloads and open-source compilers.
For data-center applications, BOLT
achieves up to 8.0\% performance speedups on top of profile-guided 
function reordering and LTO.  For the GCC and Clang compilers, our evaluation shows
that BOLT speeds up their binaries by up to 20.4\% on top of FDO and LTO, and 
up to 52.1\% if the binaries are built without FDO and LTO.
\end{abstract}

\section{Introduction}
\label{sec:intro}

\begin{figure*}[t]
\begin{center}
\includegraphics[width=1.0\linewidth]{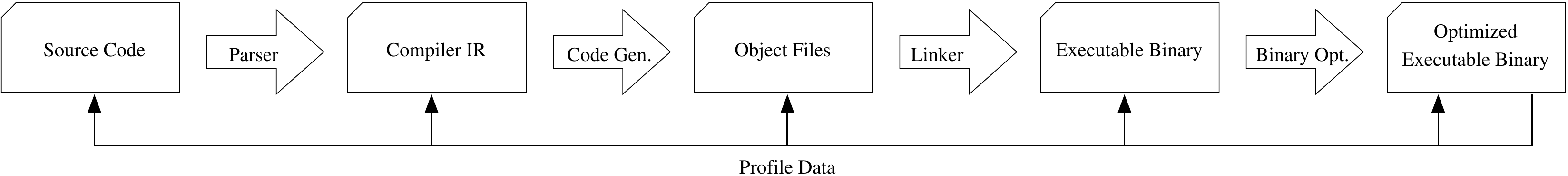}
\end{center}
\caption{Example of a compilation pipeline and the various
  alternatives to retrofit sample-base profile data.}
\label{fig:pipeline}
\end{figure*}


Given the large scale of data centers, optimizing their workloads has recently
gained a lot of interest.  Modern data-center applications tend to be
very large and complex programs.  Due to their sheer amount of code,
optimizing the code locality for these applications is very important to improve
their performance.


The large size and performance bottlenecks of data-center applications
make them good targets for \emph{feedback-driven optimizations}~(FDO), also
called \emph{profile-guided optimizations} (PGO),
particularly code layout.  At the same time, the large sizes of these
applications also impose scalability challenges to apply FDO to them.  
Instrumentation-based profilers incur significant
memory and computational performance costs, often making it impractical to gather
accurate profiles from a production system.  To simplify deployment
and increase adoption, it is desirable to have a system that can
obtain profile data for FDO from unmodified binaries running in their
normal production environments. This is possible through the use of
\emph{sample-based profiling}, which enables high-quality profiles to
be gathered with minimal operational complexity.  This is the approach
taken by tools such as Ispike~\cite{luk:04:cgo}, \linebreak
AutoFDO~\cite{chen:16:cgo}, and HFSort~\cite{ottoni:17:cgo}.  This
same principle is used as the basis of the BOLT tool presented in this
paper.


Profile data obtained via sampling can be retrofitted to multiple
points in the compilation chain.  The point into which the profile
data is used can vary from compilation time
(e.g.\ AutoFDO~\cite{chen:16:cgo}), to link time
(e.g.\ LIPO~\cite{li:10:cgo} and \linebreak HFSort~\cite{ottoni:17:cgo}), to
post-link time (e.g.\ Ispike~\cite{luk:04:cgo}).  In general, the
earlier in the compilation chain the profile information is inserted,
the larger is the potential for its impact, because more phases and
optimizations can benefit from this information.  This benefit has
motivated recent work on compile-time and link-time FDO techniques.
At the same time, post-link optimizations, which in the past were
explored by a series of proprietary tools such as
Spike~\cite{cohn:97:dec}, Etch~\cite{romer:97:usenix}, 
FDPR~\cite{henis:99:feedback}, and Ispike~\cite{luk:04:cgo}, have not attracted much
attention in recent years.  We believe that the lack of interest in
post-link optimizations is due to folklore and the intuition that this
approach is inferior because the profile data is injected very late in
the compilation chain.


In this paper, we demonstrate that the intuition described above is
incorrect.  The important insight that we leverage in this work is that,
although injecting profile data earlier in the compilation chain
enables its use by more optimizations, injecting this data later
enables more accurate use of the information for better code layout.
In fact, one of the
main challenges with AutoFDO is to map the profile data, collected at
the binary level, back to the compiler's intermediate
representations~\cite{chen:16:cgo}.  In the original compilation used
to produce the binary where the profile data is collected, many
optimizations are applied to the code by the compiler and linker
before the machine code is emitted.  In a post-link optimizer, which
operates at the binary level, this problem is much simpler, resulting
in more accurate use of the profile data.  This accuracy is
particularly important for low-level optimizations such as code layout.

We demonstrate the finding above in the context of a static binary
optimizer we built, called BOLT.  BOLT is a modern, retargetable
binary optimizer built on top of the LLVM compiler
infrastructure~\cite{lattner:04:cgo}.  Our experimental evaluation on
large real-world applications shows that BOLT can improve performance
by up to 20.41\% on top of FDO and LTO.  Furthermore, our
analysis demonstrates that this improvement is mostly due to the
improved code layout that is enabled by the more accurate usage of
sample-based profile data at the binary level.

Overall, this paper makes the following contributions:

\begin{enumerate}
\item It describes the design of a modern, open-source post-link optimizer built on
  top of the LLVM infrastructure.\footnote{{BOLT} is available at {\tt https://github.com/facebookincubator/BOLT}.}
\item It demonstrates empirically that a post-link optimizer is able
  to better utilize sample-based profiling data to improve code
  layout compared to a compiler-based approach.
\item It shows that neither compile-time, link-time, nor post-link-time
  FDO supersedes the others but, instead, they are complementary.
\end{enumerate}

This paper is organized as follows.  Section~\ref{sec:motiv} motivates
the case for using sample-based profiling and static binary
optimization to improve performance of large-scale applications.
Section~\ref{sec:arch} then describes the architecture of the BOLT
binary optimizer, followed by a description of the optimizations
that BOLT implements in Section~\ref{sec:opts} and a discussion about profiling techniques
in Section~\ref{sec:profiling}. An evaluation of BOLT
and a comparison with other techniques is presented in
Section~\ref{sec:eval}.  Finally, Section~\ref{sec:related} discusses
related work and Section~\ref{sec:conc} concludes the paper.

\section{Motivation}
\label{sec:motiv}

In this section, we motivate the post-link optimization approach used
by BOLT.

\subsection{Why sample-based profiling?}
\label{sec:motiv-sampling}

Feedback-driven optimizations~(FDO) have been proved to help increase
the impact of code optimizations in a variety of systems
(e.g.\ \cite{chen:16:cgo,dehnert:03:cgo,holzle:94:pldi,li:10:cgo,ottoni:18:pldi}).
Early developments in this area relied on \emph{instrumentation-based
  profiling}, which requires a special instrumented build of the
application to collect profile data.  This approach has two drawbacks.
First, it complicates the build process, since it requires a special
build for profile collection.  Second, instrumentation typically
incurs very significant CPU and memory overheads.  These overheads
generally render instrumented binaries inappropriate for running in
real production environments.

In order to increase the adoption of FDO in production environments,
recent work has investigated FDO-style techniques based on
\emph{sample-based
  profiling}~\cite{chen:13:ieeetoc,chen:16:cgo,ottoni:17:cgo}.
Instead of instrumentation, these techniques rely on much cheaper
sampling using hardware profile counters available in modern CPUs,
such as Intel's Last Branch Records~(LBR)~\cite{intel:manual}.  This
approach is more attractive not only because it does not require a special
build of the application, but also because the profile-collection overheads
are negligible.  By addressing the two main drawbacks of
instrumentation-based FDO techniques, sample-based profiling has
increased the adoption of FDO-style techniques in complex, real-world
production systems~\cite{chen:16:cgo,ottoni:17:cgo}.  For these same
practical reasons, we opted to use sample-based profiling in this
work.

\subsection{Why a binary optimizer?}


Sample-based profile data can be leveraged at various levels in the
compilation pipeline.  Figure~\ref{fig:pipeline} shows a generic
compilation pipeline to convert source code into machine code.  As
illustrated in Figure~\ref{fig:pipeline}, the profile data may be
injected at different program-representation levels, ranging from
source code, to the compiler's intermediate representations~(IR), to
the linker, to post-link optimizers.  In general, the designers of any
FDO tool are faced with the following trade-off.  On the one hand, injecting profile
data earlier in the pipeline allows more optimizations along the
pipeline to benefit from this data.  On the other hand, since sample-based
profile data must be collected at the binary level, the closer a level
is to this representation, the higher the accuracy with which the data
can be mapped back to this level's program representation. Therefore,
a post-link binary optimizer allows the profile data to be used with
the greatest level of accuracy.


AutoFDO~\cite{chen:16:cgo} retrofits profile data back into a
compiler's intermediate representation~(IR).  Chen et
al.~\cite{chen:13:ieeetoc} quantified the precision of the profile
data that is lost by retrofitting profile data even at a reasonably
low-level representation in the GCC compiler.  They quantified that
the profile data had 84.1\% accuracy, which they were able to improve
to 92.9\% with some techniques described in that work.

The example in Figure~\ref{fig:code} illustrates the difficulty in
mapping binary-level performance events back to a higher-level
representation.  In this example, both functions {\tt bar} and {\tt
  baz} call function {\tt foo}, which gets inlined in both callers.
Function {\tt foo} contains a conditional branch for the {\tt if}
statement on line {\tt(02)}.  For forward branches like this, on
modern processors, it is advantageous to make the most common
successor be the fall-through, which can lead to better branch
prediction and instruction-cache locality.  This means that, when {\tt
  foo} is inlined into {\tt bar}, block {\tt B1} should be placed
before {\tt B2}, but the blocks should be placed in the opposite order
when inlined into {\tt baz}.  When this program is profiled at the
binary level, two branches corresponding to the {\tt if} in line
{\tt(02)} will be profiled, one within {\tt bar} and one within {\tt
  baz}.  Assume that functions {\tt bar} and {\tt baz} execute the
same number of times at runtime.  Then, when mapping the branch
frequencies back to the source code in Figure~\ref{fig:code}, one will
conclude that the branch at line {\tt (02)} has a 50\% chance of
branching to both {\tt B1} and {\tt B2}.  And, after {\tt foo} is
inlined in both {\tt bar} and {\tt baz}, the compiler will not be able
to tell what layout is best in each case.  Notice that, although this
problem can be mitigated by injecting the profile data into a
lower-level representation after function inlining has been performed,
this does not solve the problem in case {\tt foo} is declared in a
different module than {\tt bar} and {\tt baz} because in this case
inlining cannot happen until link time.

\begin{figure}
  \scriptsize
  \begin{verbatim}
     (01)     function foo(int x) {
     (02)       if (x > 0) {
     (03)         ... // B1
     (04)       } else {
     (05)         ... // B2
     (06)       }
     (07)     }

     (08)     function bar() {
     (09)       foo(... /* > 0 */); // gets inlined
     (10)     }

     (11)     function baz() {
     (12)       foo(... /* < 0 */); // gets inlined
     (13)     }
  \end{verbatim}
\vspace{-0.3cm}
\caption{Example showing a challenge in mapping binary-level events
  back to higher-level code representations.}
\label{fig:code}
\end{figure}


Since our initial motivation for BOLT was to improve large-scale
data-center applications, where code layout plays a major role, a
post-link binary optimizer was very appealing.  Traditional
code-layout techniques are highly dependent on accurate branch
frequencies~\cite{pettis:90:pldi}, and using inaccurate profile data
can actually lead to performance degradation~\cite{chen:13:ieeetoc}.
  Nevertheless, as we mentioned earlier, feeding
profile information at a very low level prevents earlier optimizations
in the compilation pipeline from leveraging this information.
Therefore, with this approach, any optimization that we want to
benefit from the profile data needs to be applied at the binary level.
Fortunately, code layout algorithms are relatively simple and easy to
apply at the binary level.

\subsection{Why a \emph{static} binary optimizer?}


The benefits of a binary-level optimizer outlined above can be
exploited either statically or dynamically.  We opted for a static
approach for two reasons.  The first one is the simplicity of the
approach.  The second was the absence of runtime overheads.  Even
though dynamic binary optimizers have had some success in the past
(e.g.\ Dynamo~\cite{bala:00:pldi}, DynamoRIO~\cite{bruening:03:cgo},
StarDBT~\cite{wang:07:apcacsa}), these systems incur non-trivial
overheads that go against the main goal of improving the overall
performance of the target application.  In other words, these systems
need to perform really well in order to recover their overheads and
achieve a net performance win.  Unfortunately, since they need to
keep their overheads low, these systems often have to implement faster,
sub-optimal code optimization passes.  This has been a general
challenge to the adoption of dynamic binary optimizers, as they are
not suited for all applications and can easily degrade performance if
not tuned well.  The main benefit of a dynamic binary optimizer over a
static one is the ability to handle dynamically generated and
self-modifying code.

\section{Architecture}
\label{sec:arch}

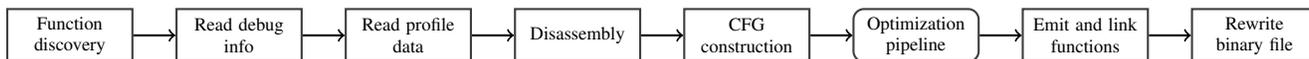
\begin{figure*}[htb]
\centering
\vspace{0.6cm}
\begin{tikzpicture}[scale=0.75]
\tikzstyle {node}= [rectangle, thick, draw=black!75,fill= black!0,
                          minimum size=7mm]
\tikzstyle {highlight_node}= [rectangle, rounded corners, thick, 
                          draw=black!75,fill= black!0,
                          minimum size=7mm]
\scriptsize

\node [node, text width=1.5cm, align=center] (n0) at (0,0) {Function discovery};
\node [node, text width=1.5cm, align=center] (n1) at (3,0) {Read debug \\ info};
\node [node, text width=1.5cm, align=center] (n2) at (6,0) {Read profile \\ data};
\node [node, text width=1.5cm, align=center] (n3) at (9,0) {Disassembly};
\node [node, text width=1.5cm, align=center] (n4) at (12,0) {CFG \\ construction};
\node [highlight_node, text width=1.5cm, align=center] (n5) at (15,0) {Optimization \\ pipeline};
\node [node, text width=1.5cm, align=center] (n6) at (18,0) {Emit and link functions};
\node [node, text width=1.5cm, align=center] (n7) at (21,0) {Rewrite binary file};

\draw [->, thick] (n0.east) to (n1.west);
\draw [->, thick] (n1.east) to (n2.west);
\draw [->, thick] (n2.east) to (n3.west);
\draw [->, thick] (n3.east) to (n4.west);
\draw [->, thick] (n4.east) to (n5.west);
\draw [->, thick] (n5.east) to (n6.west);
\draw [->, thick] (n6.east) to (n7.west);
\end{tikzpicture}
\vspace{0.25cm}
\caption{Diagram showing BOLT's binary rewriting pipeline.}
\label{fig:rewriting}
\end{figure*}

Large-scale data-center binaries may contain over 100 MB of code from
multiple source-code languages, including assembly language. In this
section, we discuss the design of the BOLT binary optimizer that we
created to operate in this scenario.

\subsection{Initial Design}

We developed BOLT by incrementally increasing its binary code
coverage. At first, BOLT was only able to optimize the code layout of
a limited set of functions. With time, code coverage gradually
increased by adding support for more complex functions.  Even today,
BOLT is still able to leave some functions in the binary untouched
while processing and optimizing others, conservatively skipping code
that violates its current assumptions.

The initial implementation targeted x86\_64 Linux ELF binaries
and relied exclusively on ELF symbol tables to guide binary content
identification. By doing that, BOLT was able to optimize code layout within existing
function boundaries. When BOLT was not able to reconstruct the control-flow 
graph of a given function with full confidence, it would just
leave the function untouched.

Due to the nature of code layout optimizations, the effective code
size may increase for a couple of reasons. First, this may happen due
to an increase in the number of branches on cold paths. Second, there
is a peculiarity of x86's conditional branch instruction, which
occupies 2 bytes if a (signed) offset to a destination fits in 8 bits
but otherwise takes 6 bytes for 32-bit offsets. Naturally, moving cold
code further away showed a tendency to increase the hot code size. If
an optimized function would not fit into the original function's
allocated space, BOLT would split the cold code and move it to a newly
created ELF segment.  Note that such function splitting was
involuntary and did not provide any extra benefit beyond allowing code
straightening optimizations as BOLT was not filling out the freed
space between the split point and the next function.

\subsection{Relocations Mode}

A second and more ambitious mode was later added to operate by changing
the position of all functions in the binary.
While multiple approaches were considered, the most obvious and
straightforward one was to rely on relocations recorded and saved by the
linker in an executable. Both BFD and Gold linkers provide such an option
(\verb|--emit-relocs|). However, even with this option, there are still
some missing pieces of information.  
An example is the relative offsets for PIC jump tables which are removed 
by the linker. Other examples are some relocations that are not
visible even to the linker, such as cross-function references for local
functions within a single compilation unit (they are processed internally
by the compiler). Therefore, in order to detect and fix such references, 
it is important to disassemble all the code correctly before trying to
rearrange the functions in the binary. Nevertheless, with relocations, the
job of gaining complete control over code re-writing became much
easier. Handling relocations gives BOLT the ability to change the order
of functions in the binary and split function bodies to further improve
code locality.

Since linkers have access to relocations, it would be possible to use
them for similar binary optimizations. However, there are multiple open-source
linkers for x86 Linux alone, and which one is being used for any
particular application depends on a number of circumstances that may
also change over time. Therefore, in order to facilitate the tool's adoption,
we opted for writing an independent post-link optimizer instead of being tied
to a specific linker.

\subsection{Rewriting Pipeline}

Analyzing an arbitrary binary and locating code and data is not
trivial. In fact, the problem of precisely disassembling machine 
code is undecidable in the general case. In practice, there is more information
than just an entry point available, and BOLT relies on correct ELF symbol table
information for code discovery. Since BOLT works with 64-bit Linux binaries,
the ABI requires an inclusion of function frame information that contains
function boundaries as well. While BOLT could have relied on this information,
it is often the case that functions written in assembly omit frame
information. Thus, we decided to employ a hybrid approach
using both symbol table and frame information when available.

Figure~\ref{fig:rewriting} shows a diagram with BOLT's rewriting steps. Function
discovery is the very first step, where function names are bound to
addresses. Later, debug information and profile data are retrieved so
that disassembly of individual functions can start.

BOLT uses the LLVM compiler infrastructure~\cite{lattner:04:cgo} to
handle disassembly and modification of binary files. There are a couple
of reasons LLVM is well suited for BOLT. First, LLVM has 
a nice modular design that enables relatively easy development of tools
based on its infrastructure. Second, LLVM supports multiple target
architectures, which allows for easily retargetable tools. To illustrate
this point, a working prototype for the ARM architecture was implemented
in less than a month. In addition to the assembler and disassembler, many
other components of LLVM proved to be useful while building BOLT. Overall,
this decision to use LLVM has worked out well. The LLVM infrastructure has
enabled a quick implementation of a robust and easily retargetable binary
optimizer.

As Figure~\ref{fig:rewriting} shows, the next step in the rewriting
pipeline is to build the control-flow graph~(CFG) representation for each
of the function.  The CFG is constructed using the \verb|MCInst| objects
provided by LLVM's Tablegen-generated disassembler. BOLT reconstructs the
control-flow information by analyzing any branch instructions encountered
during disassembly. Then, in the CFG representation, BOLT runs its optimization
pipeline, which is explained in detail in Section~\ref{sec:opts}. For BOLT,
we have added a generic annotation mechanism to \verb|MCInst| in order to
facilitate certain optimizations, e.g. as a way of recording dataflow
information. The final steps involve emitting functions and using LLVM's
run-time dynamic linker (created for the LLVM JIT systems) to resolve
references among functions and local symbols (such as basic blocks). Finally,
the binary is rewritten with the new contents while also updating ELF
structures to reflect the new sizes.


\subsection{C++ Exceptions and Debug Information}

BOLT is able to recognize DWARF~\cite{dwarf:manual}
information and update it to reflect the code modifications and relocations
performed during the rewriting pass.

Figure~\ref{fig:example} shows an example of a CFG dump demonstrating
BOLT's internal representation of the binary for the first two basic
blocks of a function with C++ exceptions and a throw statement.
The function is quite small with only five basic blocks in total, and
each basic block is free
to be relocated to another position, except the entry point.
Placeholders for DWARF \emph{Call Frame Information} (CFI) instructions are used
to annotate positions where the frame state changes (for example, when the stack
pointer advances). BOLT rebuilds all CFI for the new binary
based on these annotations so the frame unwinder works properly when
an exception is thrown.
The \verb|callq| instruction at offset \verb|0x00000010|
can throw an exception and has a designated landing pad as indicated by
a landing-pad annotation displayed next to it (\verb|handler: .LLP0; action: 1|).
The last annotation on the line indicates a source line origin for every
machine-level instruction.


\begin{figure}[t]
\scriptsize
\begin{center}
\begin{verbatim}
Binary Function "_Z11filter_onlyi" after building cfg {
  State       : CFG constructed
  Address     : 0x400ab1
  Size        : 0x2f
  Section     : .text
  LSDA        : 0x401054
  IsSimple    : 1
  IsSplit     : 0
  BB Count    : 5
  CFI Instrs  : 4
  BB Layout   : .LBB07, .LLP0, .LFT8, .Ltmp10, .Ltmp9
  Exec Count  : 104
  Profile Acc : 100.0%
}
.LBB07 (11 instructions, align : 1)
  Entry Point
  Exec Count : 104
  CFI State : 0
    00000000:   pushq   %rbp # exception4.cpp:22
    00000001:   !CFI    $0      ; OpDefCfaOffset -16
    00000001:   !CFI    $1      ; OpOffset Reg6 -16
    00000001:   movq    %rsp, %rbp # exception4.cpp:22
    00000004:   !CFI    $2      ; OpDefCfaRegister Reg6
    00000004:   subq    $0x10, %rsp # exception4.cpp:22
    00000008:   movl    %edi, -0x4(%rbp) # exception4.cpp:22
    0000000b:   movl    -0x4(%rbp), %eax # exception4.cpp:23
    0000000e:   movl    %eax, %edi # exception4.cpp:23
    00000010:   callq   _Z3fooi # handler: .LLP0; action: 1
                                # exception4.cpp:23
    00000015:   jmp     .Ltmp9 # exception4.cpp:24
  Successors: .Ltmp9 (mispreds: 0, count: 100)
  Landing Pads: .LLP0 (count: 4)
  CFI State: 3
.LLP0 (2 instructions, align : 1)
  Landing Pad
  Exec Count : 4
  CFI State : 3
  Throwers: .LBB07
    00000017:   cmpq    $-0x1, %rdx # exception4.cpp:24
    0000001b:   je      .Ltmp10 # exception4.cpp:24
  Successors: .Ltmp10 (mispreds: 0, count: 4),
              .LFT8 (inferred count: 0)
  CFI State: 3
....
\end{verbatim}
\end{center}
\caption{Partial CFG dump for a function with C++ exceptions.}
\label{fig:example}
\end{figure}

\section{Optimizations}
\label{sec:opts}

\begin{table}[t]
\begin{center}
\scriptsize
\begin{tabular}{|l|p{6cm}|}
\hline
{\bf Pass Name} & {\bf Description} \\\hline
\hline
1. strip-rep-ret & Strip \verb|repz| from \verb|repz retq| instructions used for legacy AMD processors  \\\hline
2. icf & Identical code folding  \\\hline
3. icp & Indirect call promotion \\\hline
4. peepholes & Simple peephole optimizations  \\\hline
5. inline-small & Inline small functions  \\\hline
6. simplify-ro-loads & Fetch constant data in .rodata whose address is known statically and mutate a load into a mov  \\\hline
7. icf & Identical code folding (second run)  \\\hline
8. plt & Remove indirection from PLT calls  \\\hline
9. reorder-bbs & Reorder basic blocks and split hot/cold blocks into separate sections (layout optimization)  \\\hline
10. peepholes & Simple peephole optimizations (second run)  \\\hline
11. uce & Eliminate unreachable basic blocks  \\\hline
12. fixup-branches & Fix basic block terminator instructions to match the CFG and the current layout (redone by reorder-bbs)  \\\hline
13. reorder-functions & Apply HFSort~\cite{ottoni:17:cgo} to reorder functions (layout optimization) \\\hline
14. sctc & Simplify conditional tail calls  \\\hline
15. frame-opts & Removes unnecessary caller-saved register spilling   \\\hline
16. shrink-wrapping & Moves callee-saved register spills closer to where they are needed, if profiling data shows it is better to do so  \\\hline
\end{tabular}
\end{center}
\caption{Sequence of transformations applied in BOLT's optimization pipeline.}
\label{tab2}
\end{table}

BOLT runs passes with either code transformations or analyses,
similar to a compiler. BOLT is also equipped with a dataflow-analysis
framework to feed information to passes that need it. This enables BOLT
to check register liveness at a given program point, a technique also
used by Ispike~\cite{luk:04:cgo}. 
Some passes are
architecture-independent while others are not.  In this section, we
discuss the passes applied to the Intel x86\_64 target.

Table~\ref{tab2} shows each individual BOLT optimization pass in the 
order they are applied. For example, the first line presents \verb|strip-rep-ret|
at the start of the pipeline. Notice that passes 1 and 4 are focused on leveraging
precise target architecture information to remove or mutate some instructions.
A use case of BOLT for data-center applications is to allow the user to
trade any optional choices in the instruction space in favor of I-cache
space, such as removing alignment NOPs and AMD-friendly REPZ bytes, or using shorter
versions of instructions. Our findings
show that, for large applications, it is better
to aggressively reduce I-cache occupation, except if the change incurs
D-cache overhead, since cache is one of the most constrained resources in the data-center space.
This explains BOLT's policy of discarding all NOPs after reading the input binary.
Even though compiler-generated alignment NOPs are generally useful,
the extra space required by them does not pay off and simply stripping them from the binary
provides a small but measurable performance improvement. 

BOLT features identical code folding (ICF) to complement the ICF
optimization done by the linker. An additional benefit of doing ICF at the binary
level is the ability to optimize functions that were compiled without the
\verb|-ffunction-sections| flag and functions that contain jump tables.
As a result, BOLT is able to fold more identical functions than the linkers.
We have measured the reduction of code size for the HHVM binary~\cite{adams:14:oopsla} to be about 3\%
on top of the linker's ICF pass.

Passes 3 (indirect call promotion), 5 (inline small functions), and
7 (PLT call optimization)
leverage call frequency information to either eliminate or mutate
a function call into a more performant version. We note that BOLT's function inlining
is a limited version of what compilers perform at higher levels. We expect that 
most of the inlining opportunities will be leveraged by the compiler (potentially using FDO).
The remaining inlining opportunities for BOLT are typically exposed by
more accurate profile data, BOLT's indirect-call promotion~(ICP) optimization, cross-module 
nature, or a combination of these factors.

Pass 6, simplification of load
instructions, explores a tricky tradeoff by fetching data from statically known values (in read-only sections).
In these cases, BOLT may convert loads into immediate-loading instructions, relieving
pressure from the D-cache but possibly increasing pressure on the I-cache, since the data
is now encoded in the instruction stream. BOLT's policy in this case is to abort
the promotion if the new instruction encoding is larger than the original load
instruction, even if it means avoiding an arguably more computationally expensive load instruction.
However, we found that such opportunities are not very frequent in our workloads.

Pass 9, reorder and split hot/cold basic blocks, reorders basic blocks
according to the most frequently executed paths, so the hottest successor
will most likely be a fall-though, reducing taken branches and relieving
pressure from the branch predictor unit.

Finally, pass 13 reorders the functions via the HFSort
technique~\cite{ottoni:17:cgo}.  This optimization mainly improves
I-TLB performance, but it also helps with I-cache to a smaller extent.
Combined with pass 9, these are the most effective ones in BOLT
because they directly optimize the code layout.

\section{Profiling Techniques}
\label{sec:profiling}

This section discusses pitfalls and caveats of different sample-based profiling
techniques when trying to produce accurate profiling data.

\subsection{Techniques}

In recent Intel microprocessors, LBR is a list of the last 32 taken branches.
LBRs are important for profile-guided optimizations not only because they provide
accurate counts for critical edges (which cannot be inferred even with
perfect basic block count profiling~\cite{levin:07}), but also because they 
make block-layout algorithms more resilient to bad sampling.  When evaluating 
several different sampling events to collect LBRs
for BOLT, we found that the performance impact in LBR mode is very consistent 
even for different sampling events.  We have experimented with
collecting LBR data with multiple hardware events on Intel x86, including 
\emph{retired instructions}, \emph{taken branches}, and \emph{cycles}, and also 
experimented with different levels of \emph{Precise Event Based Sampling} 
(PEBS)~\cite{intel:manual}.  In all these cases, for a workload for which 
BOLT provided a 5.4\% speedup, the performance differences were within 1\%. 
In non-LBR mode, using biased events with a non-ideal algorithm to infer edge counts
can cause as much as 5\% performance penalty when compared to LBR,
meaning it misses nearly all optimization opportunities.
An investigation showed
that non-LBR techniques can be tuned to stay under 1\% worse than LBR in this example
workload, but if LBR is available in the processor, one is better off using it to 
obtain higher and more robust performance numbers. We also evaluate this effect for HHVM
in Section~\ref{sec:eval-lbr}.

\subsection{Consequences for Block Layout}

Using LBRs, in a hypothetical worst-case biasing scenario where all samples in a
function are recorded in the
same basic block, BOLT will lay out blocks in the order of the path
that leads to this block. It is an incomplete layout that misses the
ordering of successor blocks, but it is not an invalid nor a cold
path. In contrast, when trying to infer the same edge counts with
non-LBR samples, the scenario is that of a single hot basic block
with no information about which path was taken to get to it.

In practice, even in LBR mode, many times the
collected profile is contradictory by stating that predecessors execute
many times more than its single successor, among other violations of
flow equations.\footnote{I.e., the sum of a block's input flow is equal to the sum
of its output flow.} Previous work~\cite{levin:07,novillo:14:llvm}, which includes
techniques implemented in IBM's FDPR~\cite{henis:99:fdo}, report handling the problem of
reconstructing edge counts by solving an instance of \emph{minimum cost flow}
(MCF~\cite{levin:07}), a graph network flow problem.  However, these
reports predate LBRs. LBRs only store
taken branches, so when handling very skewed data such as the cases
mentioned above, BOLT satisfies the flow equation by attributing all
surplus flow to the non-taken path that is naturally missing from the
LBR, similarly to Chen et al.~\cite{chen:13:ieeetoc}.  BOLT also benefits 
from being applied after the static compiler: to cope with uncertainty, by
putting weight on the fall-through path, it trusts the original layout
done by the static compiler.  Therefore, the program trace needs to show a
significant number of taken branches, which contradict the original 
layout done by the compiler, to convince BOLT to reorder the blocks and change 
the original fall-through path. Without LBRs, it is not possible to take
advantage of this:
algorithms start with guesses for both taken and non-taken branches without being
sure if the taken branches, those taken for granted in LBR mode, are
real or the result of bad edge-count inference.

\subsection{Consequences for Function Layout}

BOLT uses HFSort~\cite{ottoni:17:cgo} to perform function reordering based on 
a weighted call graph. If
LBRs are used, the edge weights of the call graph are directly inferred
from the branch records, which may also include function calls and returns. 
However, without LBRs, BOLT is still able to
build an incomplete call graph by looking at the direct calls in the binary and 
creating caller-callee edges with weights corresponding to the number of samples 
recorded in the blocks containing the corresponding {\tt call} instructions. 
However, this approach cannot take indirect calls into account. Even
with these limitations, we did not observe a performance penalty as severe
as using non-LBR mode for basic block reordering (Section~\ref{sec:eval-lbr})

\section{Evaluation}
\label{sec:eval}

This section evaluates BOLT in a variety of scenarios, including
Facebook server workloads and the GCC and Clang open-source
compilers.  A comparison with GCC's and Clang's PGO and LTO is also
provided in some scenarios.

The evaluation presented in this section was conducted on Linux-based
servers featuring Intel microprocessors.

\subsection{Facebook Workloads}

The impact of BOLT was measured on five binaries inside
Facebook's data centers.  The first is HHVM~\cite{adams:14:oopsla}, the
PHP virtual machine that powers the web servers at Facebook and many
other web sites, including Baidu and Wikipedia.  The second is
TAO~\cite{bronson:13:usenix}, a highly distributed, in-memory,
data-caching service used to store Facebook's social graph.  The third one is Proxygen,
which is a cluster load balancer built on top of the open-source
library with the same name~\cite{proxygen:web}.  Finally, the other
two binaries implement a service called Multifeed, which is used to
select what is shown in the Facebook News Feed.

In this evaluation, we compared the performance impact of BOLT on top
of binaries built using GCC and function reordering via
HFSort~\cite{ottoni:17:cgo}.  The HHVM binary specifically is compiled
with LTO to further enhance its performance.  Unfortunately, a
comparison with FDO and AutoFDO was not possible.  The difficulties
with FDO were the common ones outlined in
Section~\ref{sec:motiv-sampling} to deploy instrumented binaries in
these applications' normal production environments.  And we found that
AutoFDO support in the latest version of GCC available in our
environment (version 5.4.1) is not stable and caused either internal
compiler errors or runtime errors related to C++ exceptions.
Nevertheless, a direct comparison between BOLT and FDO was possible
for other applications, and the results are presented in
Section~\ref{sec:eval-compilers}.

\begin{figure}[t]
\vspace{-.5cm}
\center
\begin{tikzpicture}
[scale=1]
\tikzset{
    every pin/.style={fill=black!80,rectangle,text width=1.5cm, align=center,rounded corners=3pt,font=\scriptsize},
    small dot/.style={fill=black!80,circle,scale=0.3}
}
\pgfplotsset{width=7cm, height=1.75cm}
\begin{axis}
[ybar,
 xtick=data,
 xticklabels from table={facebook.csv}{Workload},
 bar width = 2.0ex,
 ymax=10,
 ymin=0,
 x tick label style={
          major tick length=2pt, font=\scriptsize, anchor=north, rotate=20 },
 y tick label style={
   major tick length=1pt },
 scale only axis,
 ymajorticks=true,
 ymajorgrids, 
 ylabel=\% Speedup
]
\addplot[fill=blue,error bars/.cd] 
   table[x=X, y=CPU]
{facebook.csv};
\end{axis}
\end{tikzpicture}
\vspace{-.75cm}
\caption{Performance improvements from BOLT for our set of Facebook data-center workloads.}
\label{fig:eval-facebook}
\end{figure}
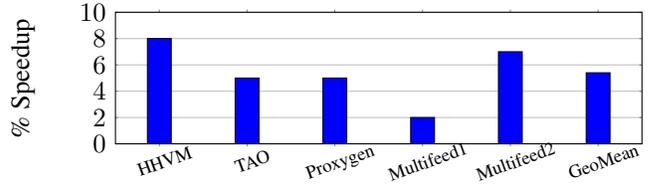

Figure~\ref{fig:eval-facebook} shows the performance results for
applying BOLT on top of HFSort for our set of Facebook
data-center workloads (and, in case of HHVM, also on top of LTO).  In all cases, BOLT's application resulted in a
speedup, with an average of 5.4\% and a maximum of 8.0\% for HHVM.
Note that HHVM, despite containing a large amount of dynamically
compiled code that is not optimized by BOLT, spends more time in
statically compiled code than in the dynamically generated code.
Among these applications, HHVM has the largest total code size, which
makes it very front-end bound and thus more amenable to the code
layout optimizations that BOLT implements.  

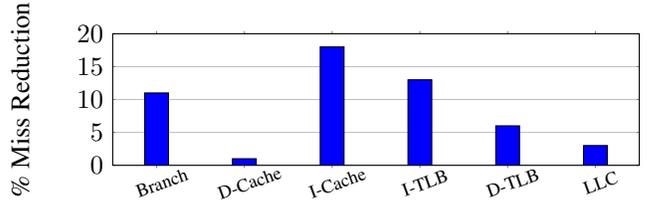
\begin{figure}[t]
\vspace{-.75cm}
\center
\begin{tikzpicture}
[scale=1]
\tikzset{
    every pin/.style={fill=black!80,rectangle,text width=1.5cm, align=center,rounded corners=3pt,font=\scriptsize},
    small dot/.style={fill=black!80,circle,scale=0.3}
}
\pgfplotsset{width=7cm, height=1.75cm}
\begin{axis}
[ybar,
 xtick=data,
 xticklabels from table={hhvm.csv}{Workload},
 bar width = 2.0ex,
 ymin=0,
 ymax=20,
 x tick label style={
          major tick length=2pt, font=\scriptsize, anchor=north, rotate=20 },
 y tick label style={
   major tick length=1pt },
 scale only axis,
 ymajorticks=true,
 ymajorgrids, 
 ylabel=\% Miss Reduction
]
\addplot[fill=blue]
   table[x=X, y=Value]
{hhvm.csv};
\end{axis}
\end{tikzpicture}
\vspace{-.75cm}
\caption{Improvements on micro-architecture metrics for HHVM.}
\label{fig:eval-hhvm}
\end{figure}


To better understand the performance benefits of BOLT, we performed a
more detailed performance analysis of HHVM.
Figure~\ref{fig:eval-hhvm} shows BOLT's improvements on important
performance metrics, including i-cache misses, i-TLB misses, branch
misses, and LLC misses. Improving branch prediction is an important
benefit from the block layout optimization done by BOLT, and for HHVM 
this metric improved by 11\%. Moreover, improving locality leads to better
metrics across the entire cache hierarchy, specially the first level
of i-cache, which exhibits 18\% reduction in misses. It is possible to see
a small improvement of 1\% in the first level of d-cache as well, due
to reordering jump tables for locality and frame optimizations. The
observed TLB improvements come from packing accessed instructions
and data into fewer pages. To better illustrate how cache and TLB locality are
improved, we present heat maps of address-space accesses in
Section~\ref{sec:eval-heatmaps}.


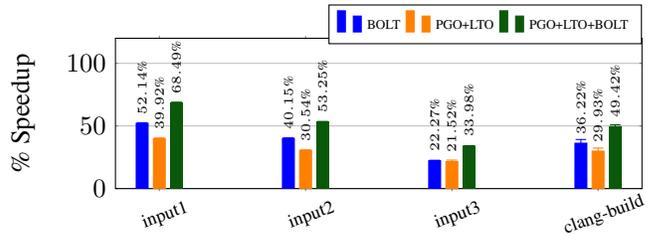
\begin{figure}[t]
\vspace{-.75cm}
\center
\begin{tikzpicture}
[scale=1]
\tikzset{
    every pin/.style={fill=black!80,rectangle,text width=1.5cm, align=center,rounded corners=3pt,font=\scriptsize},
    small dot/.style={fill=black!80,circle,scale=0.3}
}
\pgfplotsset{width=7cm, height=2cm}
\begin{axis}
[ybar,
 xtick=data,
 xticklabels from table={compilers-speedup.csv}{Workload},
 bar width = 1.0ex,
 ymin=0,
 ymax=120,
 x tick label style={yshift=-0.1cm,
          major tick length=2pt, font=\scriptsize, anchor=north, rotate=20 },
 y tick label style={
   major tick length=1pt },
 scale only axis,
 nodes near coords,
 every node near coord/.append style={
     anchor=west,
     font=\tiny,
     yshift=0.0cm,
     rotate = 90,
     color=black
   },
 nodes near coords={\pgfmathprintnumber[fixed,precision=2]{\pgfplotspointmeta}\%},
 ymajorticks=true,
 ymajorgrids, 
 legend style={at={(1,1.225)}, anchor=north east, font=\tiny, legend columns=-1},
 ylabel=\% Speedup
]
\definecolor{tempcolor1}{RGB}{10,88,10}

\addplot[blue,fill=blue,error bars/.cd,y dir=both, y explicit]
   table[x=X, y=Bolt, y error=ErrorA]
{compilers-speedup.csv};
\addlegendentry{BOLT}

\addplot[orange,fill=orange,error bars/.cd,y dir=both, y explicit]
   table[x=X, y=Fdo, y error=ErrorB]
{compilers-speedup.csv};
\addlegendentry{PGO+LTO}

\addplot[tempcolor1,fill=tempcolor1,error bars/.cd,y dir=both, y explicit]
   table[x=X, y=FdoBolt, y error=ErrorC]
{compilers-speedup.csv};
\addlegendentry{PGO+LTO+BOLT}

\end{axis}
\end{tikzpicture}
\vspace{-.75cm}
\caption{Performance improvements for Clang.}
\label{fig:eval-clang}
\end{figure}

\begin{figure}[t]
\vspace{-.5cm}
\center
\begin{tikzpicture}
[scale=1]
\tikzset{
    every pin/.style={fill=black!80,rectangle,text width=1.5cm, align=center,rounded corners=3pt,font=\scriptsize},
    small dot/.style={fill=black!80,circle,scale=0.3}
}
\pgfplotsset{width=7cm, height=2cm}
\begin{axis}
[ybar,
 xtick=data,
 xticklabels from table={gcc-speedup.csv}{Workload},
 bar width = 1.0ex,
 ymin=0,
 ymax=60,
 x tick label style={yshift=-0.1cm,
          major tick length=2pt, font=\scriptsize, anchor=north, rotate=20 },
 y tick label style={
   major tick length=1pt },
 scale only axis,
 nodes near coords,
 every node near coord/.append style={
     anchor=west,
     font=\tiny,
     yshift=0.1cm,
     rotate = 90,
     color=black
   },
 nodes near coords={\pgfmathprintnumber[fixed,precision=2]{\pgfplotspointmeta}\%},
 ymajorticks=true,
 ymajorgrids, 
 legend style={at={(1,1.225)}, anchor=north east, font=\tiny, legend columns=-1},
 ylabel=\% Speedup
]
\definecolor{tempcolor1}{RGB}{10,88,10}

\addplot[blue,fill=blue,error bars/.cd,y dir=both, y explicit]
   table[x=X, y=Bolt, y error=ErrorA]
{gcc-speedup.csv};
\addlegendentry{BOLT}

\addplot[orange,fill=orange,error bars/.cd,y dir=both, y explicit]
   table[x=X, y=Fdo, y error=ErrorB]
{gcc-speedup.csv};
\addlegendentry{PGO}

\addplot[tempcolor1,fill=tempcolor1,error bars/.cd,y dir=both, y explicit]
   table[x=X, y=FdoBolt, y error=ErrorC]
{gcc-speedup.csv};
\addlegendentry{PGO+BOLT}

\end{axis}
\end{tikzpicture}
\vspace{-.75cm}
\caption{Performance improvements for GCC. Different than Clang, we did not
use LTO due to build errors.}
\label{fig:eval-gcc}
\end{figure}
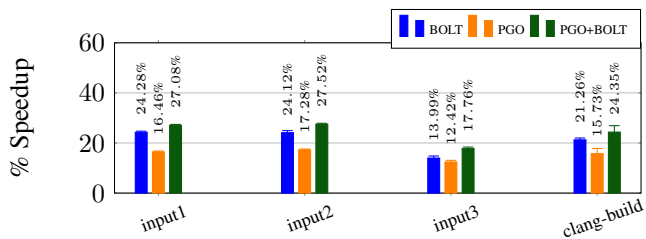

\subsection{Clang and GCC Compilers}
\label{sec:eval-compilers}

BOLT should be able to improve the performance of any
front-end bound application, not just data-center workloads. To test this theory, we ran BOLT on two open-source
compilers: Clang and GCC.

\subsubsection{Clang Setup}

For our Clang evaluation, we used the \texttt{release\_60} branch of llvm, clang, and
compiler-rt open-source repositories~\cite{llvm:web}. We built a bootstrapped release version of the
compiler first. This stage1 compiler provided a baseline for our evaluation.
We then built an instrumented version of Clang,\footnote{\texttt{-DLLVM\_BUILD\_INSTRUMENTED=ON}} and then used the instrumented
compiler to build Clang again with default options. The collected profile
data was used to do another build of Clang with LTO enabled.\footnote{\texttt{-DLLVM\_ENABLE\_LTO=Full -DLLVM\_PROFDATA\_FILE=clang.profdata}} This is
referred as \verb|PGO+LTO| in our chart.

Each of the 2 compilers was profiled with our training input, 
a full build of GCC.
We used the Linux perf utility with the option
 \texttt{record -e cycles:u -j any,u}. The profile from perf was converted using
\texttt{perf2bolt} utility into YAML format (\texttt{-w} option). Then the profile was used
to optimize the compiler binary using BOLT with the following options:

\noindent
\begin{center}
\begin{tikzpicture}%
\node[anchor=west] at (1,0.3) {
\scriptsize
\verb|-b profile.yaml -reorder-blocks=cache+|
};
\node[anchor=west] at (1,0) {
\scriptsize
\verb|-reorder-functions=hfsort+ -split-functions=3 -split-all-cold|
};
\node[anchor=west] at (1,-0.3) {
\scriptsize
\verb|-split-eh -dyno-stats -icf=1 -use-gnu-stack|
};
\end{tikzpicture}
\end{center}

The four compilers were then used to build Clang, and the overall
build time was measured for benchmarking purposes.
For all builds above we used \texttt{ninja} instead of GNU make, and for all
benchmarks we ran
them with \texttt{-j40 clang} options. We chose to build only the clang binary
(as opposed to the full build) to minimize the effect of link time on our
evaluation.

We have also selected 3 Clang/LLVM source files ranging from small to large sizes and
preprocessed those files such that they could be compiled without looking up
header dependencies. The 3 source files we used are:

\begin{itemize}
\item  input1: tools/clang/lib/CodeGen/CGVTT.cpp
\item  input2: lib/ExecutionEngine/Orc/OrcCBindings.cpp
\item  input3: lib/Target/X86/X86ISelLowering.cpp
\end{itemize}

Each of the files was then compiled with \texttt{-std=c++11 -O2} options multiple
times, and the results were recorded for benchmarking purposes.
Tests were run on a dual-node 20-core (40-core with hyperthreading) IvyBridge 
(Intel(R) Xeon(R) CPU E5-2680 v2 @ 2.80GHz) system with 32GiB RAM.

\subsubsection{GCC Setup}

For our GCC evaluation, we used version 8.2.0. First, GCC was built using the default
build process. The result of this bootstrap build was our baseline. Second, we built a
PGO version using the following configuration:

\noindent
\begin{center}
\begin{tikzpicture}%
\node[anchor=west] at (1,0.3) {
\scriptsize
\verb|--enable-linker-build-id --enable-bootstrap|
};
\node[anchor=west] at (1,0) {
\scriptsize
\verb|--enable-languages=c,c++ --with-gnu-as --with-gnu-ld|
};
\node[anchor=west] at (1,-0.3) {
\scriptsize
\verb|--disable-multilib|
};
\end{tikzpicture}
\end{center}

Afterwards, \texttt{make profiledbootstrap} was used to generate our PGO version of GCC.

Since BOLT is incompatible with GCC function splitting, we had to repeat the
above builds passing \texttt{BOOT\_CFLAGS=\'-O2 -g -fno-reorder-blocks-and-partition\'} to
the make command. The resulting compiler, ready to be BOLTed, was used to build GCC
again (our training input), this time without the bootstrap. The profile was then recorded and converted
using \texttt{perf2bolt} to YAML format, and the \texttt{cc1plus}
binary was optimized using BOLT with the same options used for Clang and later
copied over to GCC's installation directory.

All 4 different types of GCC compilers, 2 without BOLT and 2 with BOLT, were
later used to build the Clang compiler using the default configuration.

\subsubsection{Experimental Results}
Figures~\ref{fig:eval-clang} and~\ref{fig:eval-gcc} show the experimental results 
for Clang and GCC, respectively.
We observed a significant improvement on both compilers by using BOLT.
On top of GCC with PGO,
BOLT provided a 7.45\% speedup when doing a full build of Clang. On top
of Clang with LTO and PGO, a 15.0\% speedup when doing a full build of Clang.

Table~\ref{tab:stats-clang} shows some statistics reported by BOLT as
it optimizes the Clang binaries for the baseline and with PGO+LTO
applied.  These statistics are based on the input profile data.  Even
when applied on top of PGO+LTO, BOLT has a very significant impact in
many of these metrics, particularly the ones that affect code
locality.  For example, we see that BOLT reduces the number of taken
branches by 44.3\% over PGO+LTO (69.8\% over the baseline), which
significantly improves i-cache locality.

\begin{table}[t]
\centering
\footnotesize
\begin{tabular}{|l||r|r|}
\hline
Metric   &    Over Baseline     &  Over PGO+LTO \\
\hline
\hline
executed forward branches       &  -1.6\%  &  -1.0\% \\
taken forward branches          & -83.9\%  & -61.1\% \\
executed backward branches      &  +9.6\%  &  +6.0\% \\
taken backward branches         &  -9.2\%  & -21.8\% \\
executed unconditional branches & -66.6\%  & -36.3\% \\
executed instructions           &  -1.2\%  &  -0.7\% \\
total branches                  &  -7.3\%  &  -2.2\% \\
taken branches                  & -69.8\%  & -44.3\% \\
non-taken conditional branches  & +60.0\%  & +13.7\% \\
taken conditional branches      & -70.6\%  & -46.6\% \\

\hline
\end{tabular}
\caption{Statistics reported by BOLT when applied to Clang's baseline and PGO+LTO binaries.}
\label{tab:stats-clang}
\end{table}


\begin{figure*}[t]
  \centering
  \begin{subfigure}[t]{.4\linewidth}
    \centering
    \includegraphics[scale=.7]{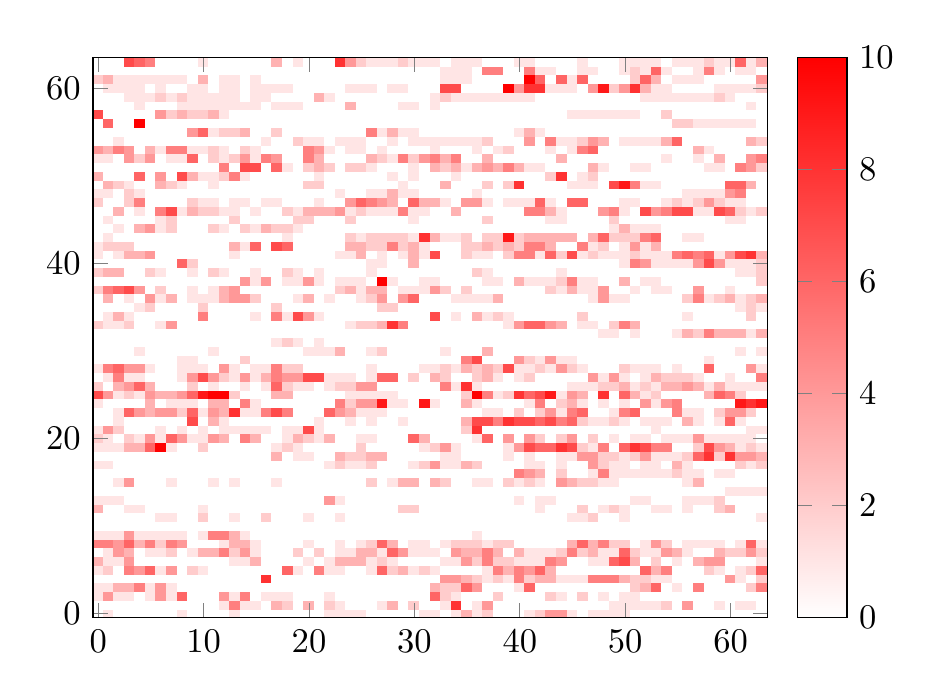}
    \caption{without BOLT}
    \label{heat1}
  \end{subfigure}
  \hspace{.075\linewidth}
  \begin{subfigure}[t]{.4\linewidth}
    \centering
    \includegraphics[scale=.7]{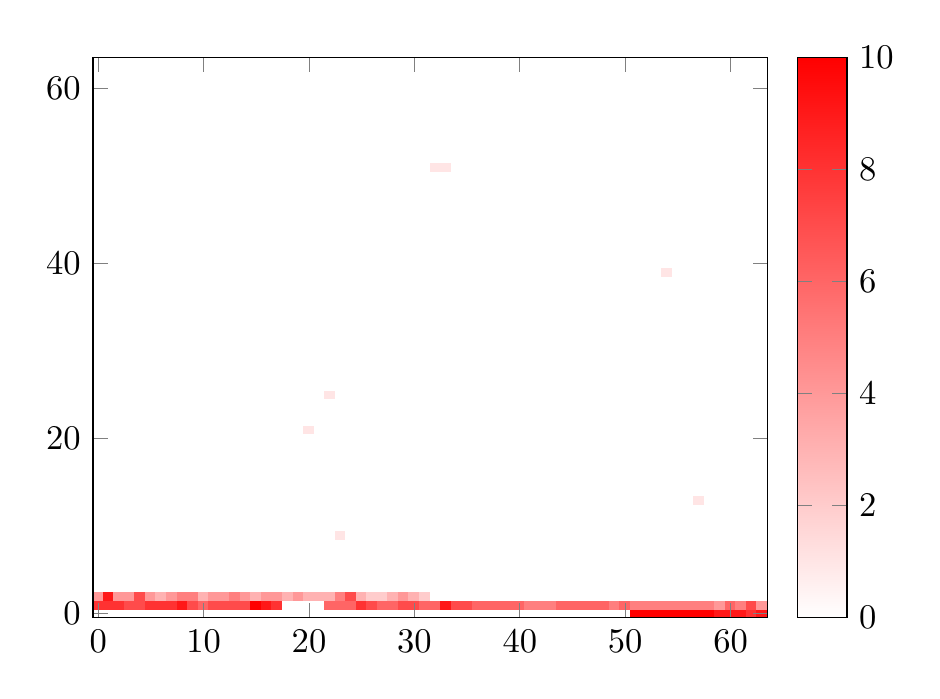}
    \caption{with BOLT}
    \label{heat2}
  \end{subfigure}
  \caption{Heat maps for instruction memory accesses of the HHVM binary, without and with BOLT. Heat is in a log scale.}
  \label{fig:heatmaps}
\end{figure*}


\begin{figure}[t]
  \scriptsize
  \begin{verbatim}
Function:
clang::Redeclarable<clang::TagDecl>::DeclLink::getNext(...)
const
  Exec Count  : 1723213

.Ltmp1100284 (4 instructions, align : 1)
  Exec Count : 1635334
  Predecessors: .Ltmp1100286, .LBB087908
    0000001d:   movq  %r12, %rbx # PointerIntPair.h:152:40
    00000020:   andq  $-0x8, %rbx # PointerIntPair.h:152:40
    00000024:   testb $0x4, %r12b # PointerUnion.h:143:9
    00000028:   je  .Ltmp1100279 # ExternalASTSource.h:462:19
  Successors: .Ltmp1100279 (mispreds: 2036, count: 1635334), 
              .LFT680413 (mispreds: 0, count: 0)

.LFT680413 (2 instructions, align : 1)
  Exec Count : 0
  Predecessors: .Ltmp1100284
    0000002a:   testq %rbx, %rbx # ExternalASTSource.h:462:19
    0000002d:   jne .Ltmp1100280 # ExternalASTSource.h:462:19
  Successors: .Ltmp1100280 (mispreds: 0, count: 0), 
              .Ltmp1100279 (mispreds: 0, count: 0)

.Ltmp1100279 (9 instructions, align : 1)
  Exec Count : 1769771
  Predecessors: .Ltmp1100284, .LFT680414, .Ltmp1100282,
                .LFT680413
    0000002f:   movq  %rbx, %rax # Redeclarable.h:140:5
    00000032:   addq  $0x28, %rsp # Redeclarable.h:140:5
    00000036:   popq  %rbx # Redeclarable.h:140:5
    00000037:   popq  %r12 # Redeclarable.h:140:5
    00000039:   popq  %r13 # Redeclarable.h:140:5
    0000003b:   popq  %r14 # Redeclarable.h:140:5
    0000003d:   popq  %r15 # Redeclarable.h:140:5
    0000003f:   popq  %rbp # Redeclarable.h:140:5
    00000040:   retq # Redeclarable.h:140:5
  \end{verbatim}
\vspace{-0.3cm}
\caption{Real example of poor code layout produced by the Clang
  compiler (compiling itself) even with PGO\@. Block {\tt .LFT680413} is
  cold ({\tt Exec Count: 0}), but it is placed between two hot blocks
  connected by a forward taken branch.}
\label{fig:poor-layout}
\end{figure}

\subsection{Analysis of Suboptimal Compiler Code Layout}
\label{sec:eval-poor-layout}

Using BOLT's {\tt -report-bad-layout} option, we inspected Clang's
binary built with LTO+PGO to identify frequently executed functions
that contain cold basic blocks interleaved with hot ones.  Combined
with options {\tt -print-debug-info} and {\tt -update-debug-sections},
this allowed us to trace the source of such blocks.  Using this
methodology, we analyzed such suboptimal code layout occurrences among
the hottest functions.  Our analysis revealed that the majority of
such cases originated from function inlining as motivated in the
example in Figure~\ref{fig:code}.  Figure~\ref{fig:poor-layout}
illustrates one of these functions at the binary level.  This function
contains 3 basic blocks, each one corresponding to source code from a
different source file. In Figure~\ref{fig:poor-layout}, the blocks are
annotated with their profile counts ({\tt Exec Count}).  The source
code corresponding to block {\tt .LFT680413} is not cold, but it is
very cold when inlined in this particular call site.  By operating at
the binary level and being guided by the profile data, BOLT can easily
identify these inefficiencies and improve the code layout.

\subsection{Heat Maps}
\label{sec:eval-heatmaps}

Figure~\ref{fig:heatmaps} shows heat maps of the instruction address
space for HHVM running with Facebook production traffic. Figure~\ref{heat1}
illustrates addresses fetched through I-cache for the regular
binary, while Figure~\ref{heat2} shows the same for HHVM processed with BOLT.

This heat map is built as a matrix of addresses. Each line has 64
blocks and the complete graph has 64 lines. The HHVM binary chosen for
this study has 148.2 MB of text size, which is fully represented
in the heat map. Each block represents 36,188 bytes and the heat map shows
how many times, on average, each byte of a block is fetched as
indicated by profiling data. For example, the line at $Y = 0$ from $X
= 0$ to $X = 63$ plots how code is being accessed in the first
2,316,032 bytes of the address space.  The average number of times a
byte is fetched is reduced by a logarithm function to help visualize the
data, so we can easily identify even code that is executed just a few
times. Completely white areas show cold basic blocks that were never
sampled during profiling, while strong red highlights the most frequently
accessed areas of instruction memory.

Figure~\ref{heat2} demonstrates how BOLT packs together hot code to use about
4 MB of space instead of the original range spanning 148.2 MB. There is
still some activity outside the dense hot area, but they are relatively 
cold. These functions were ignored by BOLT's function-reordering
pass because they have an indirect tail call. BOLT marks
functions it can fully understand as \emph{simple} as a mechanism to allow it
to operate on the binary even if it does not fully process all
functions. Those are \emph{non-simple} functions.

Indirect tail calls are
more challenging for static binary rewriters because it is difficult
to guess if the target is another function or another
basic block of the same function, which could affect the CFG.
BOLT leaves these functions untouched. This also explains a large
cold block of about 160 KB in the hot area at $Y = 1$ and $16 \leq X \leq
20$: this cold block is part of a large non-simple function whose CFG is not fully
processed by BOLT, so it is not split in the same way as other functions.

Function splitting and reordering are important to move cold basic blocks out of 
the hot area, and BOLT uses these techniques on the vast majority of
functions in the HHVM binary. The result is a tight packing of
frequently executed code as show in $Y=1$ of Figure~\ref{heat2}, which greatly benefits
I-cache and I-TLB.


%

\subsection{Importance of LBR}
\label{sec:eval-lbr}

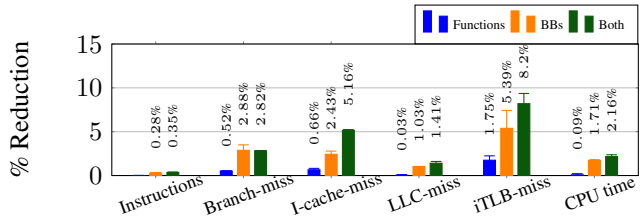
\begin{figure}[t]
\center
\begin{tikzpicture}
[scale=1]
\tikzset{
    every pin/.style={fill=black!80,rectangle,text width=1.5cm, align=center,rounded corners=3pt,font=\scriptsize},
    small dot/.style={fill=black!80,circle,scale=0.3}
}
\pgfplotsset{width=7cm, height=1.75cm}
\begin{axis}
[ybar,
 xtick=data,
 xticklabels from table={non-lbr.csv}{Workload},
 bar width = 1.0ex,
 ymin=0,
 ymax=15,
 x tick label style={
          major tick length=2pt, font=\scriptsize, anchor=north, rotate=20 },
 y tick label style={
   major tick length=1pt },
 scale only axis,
 nodes near coords,
 every node near coord/.append style={
     anchor=west,
     font=\tiny,
     yshift=0.2cm,
     rotate = 90,
     color=black
   },
 nodes near coords={\pgfmathprintnumber[fixed,precision=2]{\pgfplotspointmeta}\%},
 ymajorticks=true,
 ymajorgrids, 
 legend style={at={(1,1.3)}, anchor=north east, font=\tiny, legend columns=-1},
 ylabel=\% Reduction
]
\definecolor{tempcolor1}{RGB}{10,88,10}

\addplot[blue,fill=blue,error bars/.cd,y dir=both, y explicit]
   table[x=X, y=Function, y error=ErrorA]
{non-lbr.csv};
\addlegendentry{Functions}

\addplot[orange,fill=orange,error bars/.cd,y dir=both, y explicit]
   table[x=X, y=BB, y error=ErrorB]
{non-lbr.csv};
\addlegendentry{BBs}

\addplot[tempcolor1,fill=tempcolor1,error bars/.cd,y dir=both, y explicit]
   table[x=X, y=Both, y error=ErrorC]
{non-lbr.csv};
\addlegendentry{Both}

\end{axis}
\end{tikzpicture}
\caption{Improvements on different metrics for HHVM by using LBRs (higher is better).}
\label{fig:eval-lbr}
\end{figure}

Not all CPU vendors support a hardware mechanism to collect a trace of the last
branches, such as LBRs on Intel CPUs. We compared the impact of using them for
BOLT profile versus relying on plain samples with no such traces.

Figure~\ref{fig:eval-lbr} summarizes our evaluation on different metrics for HHVM,
in 3 different scenarios: reordering functions using HFSort, reordering basic
blocks and applying other optimizations, and with both (all optimizations on).
For example, the first data set shows that the overall reduction on instructions
executed is 0.35\% by having more accurate profiling enabled by LBRs. As
Figure~\ref{fig:eval-hhvm} shows, total CPU time improvements by using BOLT on HHVM are
about 8\%. Figure~\ref{fig:eval-lbr} shows us that using LBRs is responsible for
about 2\% of these improvements. Furthermore, the impact is more significant
for basic block layout optimizations than it is for function reordering. The reason is
because basic-block reordering requires more fine-grained profiling, at the basic-block 
level, which is harder to obtain without LBRs.

\section{Related Work}
\label{sec:related}

Binary or post-link optimizers have been extensively explored in the
past. There are two different categories for binary optimization in
general: static and dynamic, operating before program execution or
during program execution. Post-link optimizers such as BOLT are static
binary optimizers. Large platforms for prototyping and testing dynamic
binary optimizations are DynamoRIO~\cite{bruening:03:cgo} for same
host or QEMU~\cite{bellard:05:usenix} for emulation. Even though it is challenging
to overcome the overhead of the virtual machine with wins due to the
optimizations themselves, these tools can be useful in performing
dynamic binary instrumentation to analyze program execution, such as
Pin~\cite{luk:05:pldi} does, or debugging, which is the main goal of
Valgrind~\cite{nethercote:07:pldi}.

Static binary optimizers are typically focused on low-level program
optimizations, preferably using information about the precise host
that will run the program. MAO~\cite{hundt:11:cgo} is an example where
microarchitectural information is used to rewrite programs, although
it rewrites source-level assembly and not the binary itself.
Naturally, static optimizers tend to be architecture-specific.
Ispike~\cite{luk:04:cgo} is a post-link optimizer
developed by Intel to optimize for the quirks of the 
Itanium architecture. Ispike also utilizes block layout techniques
similar to BOLT, which are variations of Pettis and
Hansen~\cite{pettis:90:pldi}. However, despite supporting
architecture-specific passes, BOLT was built on top of
LLVM~\cite{lattner:04:cgo} to enable it to be easily ported to
other architectures.  
Ottoni and Maher~\cite{ottoni:17:cgo} present an enhanced
function-reordering technique based on a dynamic call graph. BOLT implements
the same algorithm in one of its passes.

Profile information is most commonly used to augment the compiler to
optimize code based on run-time information, such as done by
AutoFDO~\cite{chen:16:cgo}. The latter has also been studied in the
context of data-center applications, like BOLT.  Even though there is
some expected overlap in gains between AutoFDO and BOLT, since both
tools perform layout, in this paper we show that the gains with FDO in
general (not just AutoFDO) and BOLT can be complimentary and both
tools can be used together to obtain maximum performance.

\section{Conclusion}
\label{sec:conc}

The complexity of data-center applications often results in large
binaries that tend to exhibit poor CPU performance due to significant
pressure on multiple important hardware structures, including caches,
TLBs, and branch predictors. To tackle the challenge of improving
performance of such applications, we created a post-link optimizer,
called BOLT, which is built on top of the LLVM infrastructure.  The
main goal of BOLT is to reorganize the applications' code to reduce
the pressure that they impose on those important hardware
structures. BOLT achieves this goal with a series of optimizations,
with particular focus on code layout.  A key insight of this paper is
that a post-link optimizer is in a privileged position to perform
these optimizations based on profiling, even beyond than what a
compiler can achieve.

We tested our assumptions in Facebook data-center
applications and obtained improvements ranging from 2\% to 8\%. Unlike
profile-guided static compilers, BOLT does not need to retrofit
profiling data back to source code, making the profile more
accurate. Nevertheless, a post-link optimizer has fewer
optimizations than a compiler. We show that the strengths of both
strategies combine instead of purely overlapping, indicating that
using both approaches leads to the highest efficiency for large, front-end
bound applications. To show this, we measure the performance
improvements on two open-source compilers, GCC and Clang, featuring
large code bases dependent on the instruction cache
performance. Overall, BOLT achieves 15\% performance improvement for
Clang on top of LTO and FDO.

\begin{acks}                            
  We would like to thank Gabriel Poesia and Theodoros Kasampalis for
their work on BOLT during their internships at Facebook. We would
also like to thank Sergey Pupyrev for his work on improving the basic
block layout algorithms used by BOLT.
\end{acks}

\bibliographystyle{ieeetr}



\end{document}